%% file: main.tex
\def\year{2021}\relax
\documentclass[letterpaper]{article} %
\usepackage{aaai21}  %
\usepackage{times}  %
\usepackage{helvet} %
\usepackage{courier}  %
\usepackage[hyphens]{url}  %
\usepackage{graphicx} %
\usepackage{natbib}  %
\usepackage{caption} %
\usepackage{xcolor}
\usepackage{amsmath}
\usepackage{amssymb}
\usepackage{url}
\usepackage{booktabs}
\usepackage{multirow}
\usepackage{graphbox}
\usepackage{cite}
\usepackage{soul}
\usepackage{subcaption}
\definecolor{ForestGreen}{RGB}{34,139,34}

\DeclareMathOperator*{\argmin}{arg\,min}

\title{Exploring the Boundaries of Ambient Awareness in Twitter}
\author{
    Pablo S\'anchez-Mart\'in\textsuperscript{\rm 1 3} \hspace{0.5cm}
    Sonja Utz\textsuperscript{\rm 2}  \hspace{0.5cm}
    Isabel Valera\textsuperscript{\rm 3 4}
    \\
}
\affiliations{
    \textsuperscript{\rm 1}Max Planck Institute for Intelligent Systems, T\"{u}bingen, Germany\\
    \textsuperscript{\rm 2}Leibniz-Institut f\"{u}r Wissensmedien, T\"{u}bingen, Germany\\
    \textsuperscript{\rm 3}Saarland University, Saarbr\"{u}cken, Germany\\
    \textsuperscript{\rm 4}Max Planck Institute for Software Systems, Saarbr\"{u}cken, Germany\\
     psanchez@tue.mpg.de, s.utz@imw-tuebingen.de, ivalera@cs.uni-saarland.de
}

\begin{document}

\input{sec/commands}

\maketitle
\frenchspacing

\begin{abstract}

\emph{Ambient awareness} refers to the ability of social media users to obtain knowledge about who knows what (i.e., users' expertise) in their network, by simply being exposed to other users' content (e.g, tweets on Twitter).
Previous work, based on user surveys, reveals that individuals self-report ambient awareness only for parts of their networks. However, it is unclear whether it is their limited cognitive capacity or the limited exposure to diagnostic tweets (i.e., online content) that prevents people from developing ambient awareness for their complete network.
In this work, we focus on \emph{in-wall ambient awareness} (IWAA) in Twitter and conduct a two-step data-driven analysis, %
that allows us to explore to which extent IWAA is likely, or even possible.
First, we rely on reactions (e.g., likes), as strong evidence of users being aware of experts in Twitter. 
Unfortunately, such strong evidence can be only measured for active users, which represent the minority in the network. 
Thus to study the boundaries of IWAA to a larger extent, in the second part of our analysis we instead focus on the passive exposure to content generated by other users---which we refer to as \emph{in-wall visibility}. %
This analysis shows that (in line with \citet{levordashka2016ambient}) only for a subset of users IWAA is plausible, while for the majority it is unlikely, if even possible, to develop IWAA. 
We hope that our methodology paves the way for the emergence of data-driven approaches for the study of ambient awareness.

\end{abstract}

\input{sec/01-introduction.tex}

\input{sec/02-related_work}

\input{sec/03-problem_statement}

\input{sec/04-dataset}

\input{sec/05-iwaa_interaction}

\input{sec/06-iwaa_visibility}

\input{sec/07-discussion}

\appendix
\input{appendix/app_dataset}

\input{appendix/app_clustering}

\input{appendix/app_visibility}

{\small
\bibliography{bibliography}
}

\end{document}

%% file: sec/commands.tex
 \newcommand{\Xb}{{\boldsymbol{\X}}}
 \newcommand{\Zb}{{\boldsymbol{\Z}}}
 \newcommand{\xb}{{\boldsymbol{x}}}
  \newcommand{\eb}{{\boldsymbol{e}}}
 \newcommand{\zb}{{\boldsymbol{z}}}
 \newcommand{\betab}{{\boldsymbol{\beta}}}
  \newcommand{\thetab}{{\boldsymbol{\theta}}}
 \newcommand{\phib}{{\boldsymbol{\phi}}}
 \newcommand{\gammab}{{\boldsymbol{\gamma}}}
  \newcommand{\mub}{{\boldsymbol{\mu}}}
  \newcommand{\bbold}{{\boldsymbol{b}}}

 \newcommand{\pib}{{\boldsymbol{\pi}}}
 \newcommand{\omegab}{{\boldsymbol{\omega}}}
 \newcommand{\etab}{{\boldsymbol{\eta}}}
 \newcommand{\xtildeb}{{\widetilde{\xb}}}
  \newcommand{\wtilde}{{\widetilde{w}}}
 \newcommand{\xlineb}{{\overline{\xb}}}
 \newcommand{\etatildeb}{{{\widetilde{\etab}}}} 
 
  \newcommand{\Yhat}{{\hat{Y}}}
    \newcommand{\Shat}{{\hat{S}}}
  
   \newcommand{\Kcal}{ {  \mathcal{K} } }
   \newcommand{\Vcal}{ {  \mathcal{V} } }
     \newcommand{\Ucal}{ {  \mathcal{U} } }
     \newcommand{\Mcal}{ {  \mathcal{M} } }
\newcommand{\Bcal}{ {  \mathcal{B} } }
\newcommand{\Scal}{ {  \mathcal{S} } }
\newcommand{\Fcal}{ {  \mathcal{F} } }
\newcommand{\Ecal}{ {  \mathcal{E} } }
 \newcommand{\bigO}[1]{\mathcal{O}\left( #1\right) }
 
 \newcommand{\bydef}{{:=}}
  \newcommand{\detm}{{\operatorname{det} }}

 \newcommand{\g}[2][]{g_{#1}\left(#2\right)} %
  \newcommand{\f}[2][]{f_{#1}\left(#2\right)} %
    \newcommand{\h}[2][]{h_{#1}\left(#2\right)} %
 \newcommand{\p}[2][]{p_{#1}\left(#2\right)} %
  \newcommand{\q}[2][]{q_{#1}\left(#2\right)} %
  \newcommand{\sqf}[2][]{\Pi\left(#1; #2\right)} %
 \newcommand{\realn}{{\mathbb{R}}} %
 \newcommand{\nonegn}{{\realn_{\geq 0}}} %
 
 \newcommand{\erf}[1]{\text{erf}(#1)}
 
\newcommand{\liminfi}[2][]{\underset{#1 \rightarrow \infty}{\lim } #2} %
 \newcommand{\ind}[1]{\mathbb{I}\left[ #1 \right]}
 
\newcommand{\mbbE}{\mathbb{E}}
\newcommand{\mbbP}{\mathbb{P}}
\newcommand{\mbbR}{\mathbb{R}}

\newcommand{\mcK}{\mathcal{K}}

\newcommand{\IWAE}[1][]{\text{IWAE}_{#1}}
\newcommand{\ELBO}{\text{ELBO}}
\newcommand{\KL}{\text{KL}}
\newcommand{\SUMO}{\text{SUMO}}

\newcommand{\name}{\text{TULE}}

\newcommand{\E}[2][]{\mathbb{E}_{#1}\left[#2\right]}

 \newcommand{\mse}[2][]{\text{MSE}_{#1}\left( #2 \right)}
 \newcommand{\msehat}[2][]{\hat{\text{MSE}}_{#1}\left( #2 \right)}

\newcommand{\var}[2][]{\mathbb{V}_{#1}\left( #2 \right)}
\newcommand{\bias}[1]{\mathbb{B}\left( #1 \right)}
\newcommand{\Ncal}{ {  \mathcal{N} } }
\newcommand{\Geom}[1]{ {  \text{Geometric}\left( #1 \right) } }
\newcommand{\Expon}[1]{ {  \text{Exp}\left( #1 \right) } }

\newcommand{\note}[2][]{{\color{red} (#1: #2)}}
\newcommand{\needcite}{ {\color{red}[cite]}  }
\newcommand{\ask}[1][]{ {\color{red}#1}  }
\newcommand{\iv}[2][]{{\color{violet} (IV: #2)}}
\newcommand{\isa}[1]{\textcolor{violet}{#1}}

\newcommand{\PS}[1]{\textcolor{ForestGreen}{#1}}
\newcommand{\PScom}[1]{[\textcolor{ForestGreen}{PS: #1}]}

\newcommand{\IV}[1]{\textcolor{violet}{#1}}
\newcommand{\IVcom}[1]{[\textcolor{violet}{IV: #1}]}

\newcommand{\SU}[1]{\textcolor{orange}{#1}}
\newcommand{\SUcom}[1]{[\textcolor{orange}{SU: #1}]}

\newtheorem{thm}{Theorem}

\newtheorem{defn}[thm]{Definition} %
\newtheorem{exmp}[thm]{Example} %
\newtheorem{cor}[thm]{Corollary} %

\newcommand{\be}{\begin{equation}}
\newcommand{\ee}{\end{equation}} 
\newcommand{\bea}{\begin{eqnarray}}
\newcommand{\eea}{\end{eqnarray}}
\newcommand{\avg}[2][]{\langle #2 \rangle_{#1}}
\let\setunion=\ccup
\newcommand{\ccup}[1]{\left\{#1\right\}}
\let\setunion=\bup
\let\setunion=\rup
\newcommand{\bup}[1]{\left(#1\right)}
\newcommand{\rup}[1]{\left[#1\right]}

\newcommand{\M}{\mathcal{M}}
\newcommand{\ity}[2][]{\lambda_{#1}\left(#2\right)} %
\newcommand{\pres}[2][]{o_{#1}\left(#2\right)} %
\newcommand{\lhry}[2][]{{H^l_{#1}\left(#2\right)}} %

\newcommand{\ipres}[2][]{O_{#1}\left(#2\right)} %
\newcommand{\bity}[2][]{\beta_{#1}^*\left(#2\right)} %
\newcommand{\hry}[2][]{{H_{#1}\left(#2\right)}} %

\newcommand{\itywall}[2][]{\mu_{#1}\left(#2\right)} %
\newcommand{\hrywall}[2][]{H_{#1}^{W}\left(#2\right)} %
\newcommand{\hrywallu}[3][]{H_{#1}^{#2}\left(#3\right)} %

\newcommand{\hryrep}[2][]{H_{#1}^{A}\left(#2\right)} %

\newcommand{\vis}[2][]{\alpha_{#1}\left(#2\right)} %

\newcommand{\IRT}[1]{I_{\text{#1}}^{\text{r}}} %
\newcommand{\ERT}{E^{\text{r}}} %
\newcommand{\IRTs}{I_{\text{s}}^{\text{r}}} %
\newcommand{\IRTe}{I_{\text{e}}^{\text{r}}} %

\newcommand{\Ilik}[1]{I_{\text{#1}}^{\text{l}}} %
\newcommand{\Elik}{E^{\text{l}}} %
\newcommand{\Iliks}{I_{\text{s}}^{\text{l}}} %
\newcommand{\Ilike}{I_{\text{e}}^{\text{l}}} %

\newcommand{\Irep}[1]{I_{\text{#1}}^{\text{a}}} %
\newcommand{\Erep}{E^{\text{a}}} %
\newcommand{\Ireps}{I_{\text{s}}^{\text{a}}} %
\newcommand{\Irepe}{I_{\text{e}}^{\text{a}}} %

\newcommand{\Ix}{I_{\text{es}}^{\text{x}}} %
\newcommand{\Ex}{E^{\text{x}}} %
\newcommand{\Ixs}{I_{\text{s}}^{\text{x}}} %
\newcommand{\Ixe}{I_{\text{e}}^{\text{x}}} %

\newcommand{\Vis}{V_{\text{es}}} %
\newcommand{\VisLBs}{V^{\text{LB}}} %
\newcommand{\VisLB}{V_{\text{es}}^{\text{LB}}} %
\newcommand{\VisUB}{V_{\text{es}}^{\text{UB}}} %
\newcommand{\VisUBs}{V^{\text{UB}}} %

\newcommand{\Vish}{\hat{V}_{\text{es}}} %
\newcommand{\VishLB}{\hat{V}_{\text{es}}^{\text{LB}}} %
\newcommand{\VishUB}{\hat{V}_{\text{es}}^{\text{UB}}} %

\newcommand{\EVis}{E^{\text{v}}} %
\newcommand{\Vise}{V_{\text{e}}} %
\newcommand{\Viss}{V_{\text{s}}} %

\newcommand{\Ifol}[1]{I_{\text{#1}}^{\text{f}}} %

\newcommand{\activity}[1]{\mathcal{A}_{#1}} %
\newcommand{\activityt}[1]{\mathcal{A}^t_{#1}} %
\newcommand{\activityp}[1]{\mathcal{A}^p_{#1}} %
\newcommand{\activityr}[1]{\mathcal{A}^r_{#1}} %
\newcommand{\activitya}[1]{\mathcal{A}^a_{#1}} %
\newcommand{\activityl}[1]{\mathcal{A}^l_{#1}} %

\newcommand{\wall}[1]{\mathcal{W}_{#1}} %

\newcommand{\friends}[1]{\mathcal{B}_{#1}} %
\newcommand{\followers}[1]{\mathcal{F}_{#1}} %

\newcommand{\List}{List} %
\newcommand{\listT}[1]{L_{#1}} %
\newcommand{\listtime}{t_l} %

\newcommand{\sumI}[1]{I_{#1}} %

\newcommand{\eset}[1]{\mathcal{E}_{#1}} %

\newcommand{\graphicalone}[1]{\vspace*{\fill}\begin{center}\includegraphics{#1}\end{center}\vspace*{\fill}}

%% file: sec/01-introduction.tex
\section{Introduction}
\label{sec: introduction}
Nowadays, online social networks (OSNs) have become an indispensable part of many people's daily life. It became normal that people check their social media feeds several times a day \cite{gerlach2020constant,reinecke2018permanently}, and, consequently, are exposed to a vast stream of messages--e.g., tweets in Twitter. Browsing feeds has often been considered a waste of time or a form of addiction \cite{gerlach2020constant}. On the contrary, it has also been shown that even superficial skimming of seemingly mundane online updates helps people to develop awareness about \emph{who knows what/whom} within their OSNs.
 This form of peripheral online social awareness is defined by social scientists as \emph{ambient awareness} \cite{levordashka2016ambient,thompson2008brave}, and it happens in the absence of extensive one-to-one communication and without the awareness being the ultimate goal. The benefits of ambient awareness have mainly been studied in the context of organizational knowledge management.
Early knowledge management took an engineering approach through the use of databases but the time and effort required to store/search information turned out as a major barrier. 
Instead, the emergent approaches focused on connecting employees with the respective experts in the company \cite{van2009managing}. Recently, big hopes are set into enterprise social media because employees now could learn who-knows-what as an effortless by-product of skimming updates \cite{kane2017evolutionary,leonardi2015ambient}. Nonetheless, studies in organizations are challenged by the fact that it is difficult to determine how much of this “ambient” awareness stems indeed from skimming social media updates instead of offline contact with colleagues.

To overcome this difficulty, \citet{levordashka2016ambient} studied ambient awareness on Twitter, a platform where it is common to follow strangers \cite{utz2016linkedin}. They observed that most respondents reported ambient awareness, albeit only for parts of their network. 
Yet, it is unclear whether this limited development of ambient awareness is mainly due to the finite cognitive capacity of the users, or the fact that many Twitter users do not post enough to allow others to infer their expertise. 
Twitter can be used for many different purposes \cite{kwak2010twitter} and talking about one’s work and expertise is just one possible motivation \cite{java2007we}; it could thus be that only a small proportion of users post messages that allow to infer others' expertise. However, prior work has found that not only users of enterprise social media and professional social network sites (e.g., LinkedIn) but also Twitter users report higher professional informational benefits than non-users \cite{utz2016linkedin, leonardi2015ambient}, which might be driven by the development of ambient awareness.

In this work,  we approach ambient awareness in Twitter from a data-driven perspective. We focus our study on Twitter since i) it prevails as one of the leading platform for information exchange, and ii) users usually follow large number of  distant friends or even strangers \cite{utz2016linkedin},  which makes it a suitable OSN for the study of ambient awareness.
We propose to quantify  ``awareness signals'', namely reactions and exposure to content, in order to study the limits of ambient awareness and 
eventually obtain insights about
its plausibility or impossibility for each Twitter user.
To this end, we first define \emph{in-wall ambient awareness} (IWAA) as a particular type of ambient awareness that arises from the content of a user's wall\footnote{The wall (a.k.a. home timeline) is a stream of tweets \url{https://help.twitter.com/en/using-twitter/twitter-timeline}}--where Twitter users spend ${\sim}82\%$ of their sessions~\cite{meier2017using}.
Moreover, we rely on previous work \cite{bhattacharya2014deep} that makes use of the \List\ feature of Twitter to identify topical experts, as well as information seekers, i.e., users who created a \List\ and hence are likely to seek information on the corresponding topic. %
We assume that the creation time of a seeker's \List\ informs us of the instant when a seeker confirms their awareness of an expert. 
With this information, we quantify to which extent the seeker was exposed to the expert’s content prior to the List creation, which we use to assess the plausibility of IWAA for each of the seekers.

We divide our analysis into two steps, which in turn make use of different types of data: the first part focuses on users' interactions, while the latter measures passive exposure.
More specifically, we first study IWAA by quantifying the reactions--e.g., retweets, replies, and likes--of seekers to experts’ content. While in this analysis we find out three different seeker profiles, only $\sim$10\% of the seekers exhibit reactions to the experts’ content. In other words, $\sim$90\% of the seekers do not provide strong evidence of exposure to the tweets of the experts. This is, however, not a necessary precondition of ambient awareness; expertise inferences should also occur just by skimming updates in a completely passive manner. 
This motivates our second part of the analysis, where we do not require seekers to react to experts’ content--the larger share of Twitter users are ``lurkers'' since $10\%$ of Twitter users create $80\%$ of the tweets \cite{wojcik2019sizing}--to quantify the plausibility of IWAA. In this second analysis, we focus on exposure to experts’ content: we measure the time the content of an expert is exposed in the seeker’s wall to quantify \emph{in-wall visibility}. Notice that if the expert’s content is not visible in the seeker’s wall, then IWAA is not possible at all. Since it is not possible to compute the true visibility, we compute an upper and lower bound of it. 
This analysis unveils that, for the ``lurker'' users, i) IWAA is plausible only for $\sim$10\% of the seekers; while
ii) for over $\sim$60\% of the seekers it is unlikely, if even possible, to develop IWAA, and thus experts in their List have been discovered by the seeker outside the wall.

These results form an important contribution to work on ambient awareness as they indicate that it might not be limited cognitive capacity, but limited exposure to diagnostic tweets what prevents people from inferring expertise for all of their network members. 

\vspace{-0.4cm}
\paragraph{Notation}  We denote random variables with capital letters $A$ and specific realizations of a random variable in lower-case, i.e., $A=a$. We denote the expectations with respect to a random variable $A$ as $\avg[A]{\cdot}=\E[A]{\cdot}$. 
We denote sets/sequences with calligraphic letters, i.e.,  $\activity{}$, and their cardinality (number of elements) as $|\activity{}|$. We use  $\activity{}(t_1, t_2)$ to refer to the elements of a sequence within an interval $[t_1, t_2)$, and $\activity{}(t_2) = \activity{}(0, t_2)$.

%% file: sec/02-related_work.tex
\section{Related Work}
\label{sec:related_work}

\paragraph{Ambient awareness} is defined as ``awareness of others, arising from the frequent reception of fragmented personal information, such as status updates and various digital footprints, while browsing social media'' \cite{levordashka2016ambient}. In the context of organizational knowledge management and enterprise social media, the focus is more specifically on awareness of the expertise of network members \cite{leonardi2015ambient}. 
Field studies in organizations have shown that using enterprise social media is related to better knowledge management \cite{leonardi2014social,leonardi2015ambient,leonardi2015social,zhao2020features}. From these studies, it remains open how “ambient” this process indeed is. The idea is that ambient awareness develops without a specific goal to do so, even without deliberate attention and rather as a by-product of social media use. In organizations, it cannot be excluded that people also talk on other media or face-to-face to their colleagues. To examine the "ambient" nature of the awareness, previous work studied the ambient awareness of Twitter users for people they know only via Twitter \cite{levordashka2016ambient}. Although most participants reported experiencing ambient awareness, they usually did so for ``some'' of the members of their Twitter network.

The development of ambient awareness can be explained by spontaneous inferences, which have been largely studied by social psychologists \cite{newman1993individualists,uleman1996line}. 
A key question has been whether people can make inferences about others’ personality traits automatically. Automatic processes occur without awareness, intention and cognitive effort and they are hard to control \cite{bargh1994four}.
 A classical paradigm to study these processes is the false recognition paradigm. People read sentences such as ``Monica got an A in the math test'' and have to indicate quickly with a key-press whether words such as ``test'', ``intelligent'', or ``table'' have been in the sentence. These studies show that people more often make mistakes when traits than can be inferred from the sentence (in this example, ``intelligent'') are presented. 
 Later work has shown that spontaneous inference occur also for goals and roles, and that they are also tied to faces (vs. names) \cite{chen2014spontaneous,moskowitz2016spontaneous,todorov2002spontaneous,todorov2003efficiency}. \citet{utz2017knowledge} adapted the false recognition paradigm to the context of social media and showed in experiments with construed material that people also make trait and expertise inferences on social media. Up to the best of our knowledge, the present work constitutes the first data-driven analysis of ambient awareness in Twitter.
 
 \vspace{-0.2cm}
 \paragraph{Topical group identification} For our analysis, we first need to find topical groups (i.e., groups of users that share interest in a specific topic) and also identify the role of the users inside the group (i.e., seeker of information or expert). In the context of Twitter,  diverse  methodologies have been applied to solve this problem. 
Some approaches leverage \emph{content generated} \cite{blei2003lda, bi2014scalable, pal2011identifying} ---i.e. tweets--- and/or the \emph{network structure} \cite{page1999pagerank, weng2010twitterrank}.
These approaches however fail at accounting for the fact that usually following-follower relationships are bond-based instead of identity-based \cite{java2007we}. 
As a consequence, the inferred communities do not clearly distinguish between  popularity and expertise; and more importantly, do not represent well-defined topics, which plays an imperative role in the experts' discovery~\cite{yang2015defining}.

Alternative approaches rely instead on crowd-knowledge \cite{ghosh2012cognos, sharma2012inferring}, following the axiom stating that the average over many noisy voices leads to better outcomes than individual experienced judgements. In Twitter, these approaches  mostly rely on meta-data of the \List\ feature, and tend to outperform the other approaches in terms of both cluster cohesion and coverage of topics and users.
For our analysis, we will rely on \citet{bhattacharya2014deep} as the starting point. \citet{bhattacharya2014deep} show that the name and description of the \List\ feature in Twitter offer semantic cues to the expertise of the users included in the \List. Their methodology allows to identify the creators of the Lists as users interested in a topic--i.e. \emph{seekers} of information--and users included in  the \List s as potential \emph{experts}. 
Specifically, they consider a user to be an expert on a topic, if he is listed at least 10 times on \List s associated to the corresponding subject--they set this threshold based on previous findings \cite{ghosh2012cognos}. Additionally, the creation time of the \List\ provides the time instant from which the seeker confirms awareness of the experts.

%% file: sec/03-problem_statement.tex
\section{In-wall ambient awareness}
\label{sec:problem_statement}

Twitter is an OSN that offers its users different ways to consume other users' content, also known as timelines: Explore, Profiles, List, and Home (a.k.a. wall) timelines. They differ in how the information is presented---the sections in which the timelines are divided--- and the users whose content appear. 
While ambient awareness can happen in any of these places, previous work states that the main usage of Twitter, in fact 82\% of the sessions, corresponds to browsing the wall~\cite{meier2017using}. 
Motivated by this, we focus our study of ambient awareness to the wall, and thus, of \emph{in-wall ambient awareness} (IWAA).
Specifically, we ask: \emph{is it possible that the seekers  get to know who are experts on a topic, and thus, add them to a List, due to their usage of Twitter?} 

\vspace{-0.4cm}
\paragraph{Problem statement} Consider that seeker $s$  has created a List on a particular topic (e.g. Math) at time $t_l$. Within the members of the List there is another user $e$, who is an expert on the topic (we remind that we consider a user to be an expert on a topic, only if he is listed in at least 10 Lists on that topic).
Here we aim to quantify the exposure of the seeker to the content of the expert, and thus study the plausibility of IWAA, in a time interval previous to the creation of the List, that is  $[t_l - T, t_l]$. To this end, we consider the following two types of phenomena occurring within the wall: %
\begin{itemize}
    \item \textbf{Reactions to content: } The seeker reacts to one or more tweets of the expert. This reaction can happen in different ways: i) $s$ replies to the content of $e$; ii) $s$ acknowledges the content of $e$ through  a like or a retweet. The first situation, while being strong evidence of awareness, involves content generation which implies cognitive effort, and thus is unclear how ``ambient'' it is. In contrast, the second type of reaction is essentially cognitive-effortless and,  therefore, provides better evidence of IWAA. We refer to this type of interactions as \emph{effortless reactions}.
    
    \item \textbf{Exposure to  content: } The content of the expert is \emph{visible} in the wall of the seeker. For this condition to happen either the seeker follows the expert or one of his friends retweet his content. In contrast to the content reactions, it does not require any action (i.e., effort) from the seeker.
\end{itemize}

In order to quantify the seeker's reactions to the expert's content, 
we need to reconstruct the online activity of the seeker $\activity{s}= \{\activityp{s} \cup \activityl{s} \}$, which accounts for both posts  $\activityp{s}$
and likes $\activityl{s}$.  The sequence of posts $\activityp{s}$ can be further divided in tweets $\activityt{s}$, retweets $\activityr{s}$ and replies/answers to other tweets $\activitya{s}$. 
Furthermore, to measure the expert's content visibility  in the seeker's wall, we need to reconstruct the seeker's wall, $\wall{s} = \bigcup_{ b \in \friends{s}}\activityt{b} \cup \activityr{b}$, where $\friends{s}$ denotes the set of of friends (a.k.a. followees) of the seeker. Notice that the replies do not appear in the wall. 
Also, we remark here that the expert competes with the rest of the seeker's friends for the seeker's attention. 
As before, we denote the expert's online activity by $\activity{e}$, which can be partitioned likewise, and his set of followers by $\followers{e}$. 
\\

Given all these data we aim to explore and quantify how much has the seeker been \textit{"exposed and acknowledged the expert's content"}, since this will provide evidence about the plausibility or impossibility of IWAA. 
As a remark, note the study of ambient awareness, and in particular IWAA, involves a local problem: we focus on online behaviour of one expert and one seeker. This differs with the topical expert identification where the problem is posed as a global problem that involves all of the network. 

\vspace{-0.4cm}
\paragraph{Main assumptions of the analysis} For the analysis carried out in the following sections, we make the following assumptions:
i) the creation time of a List $t_l$ points out the instant in which the seeker confirms awareness of the expert, i.e.  all the members of a List are added at creation time. In this way, the "ambient" cannot be due to the List timeline; 
ii) the network, i.e., the  follower \& following connections, are assumed to be static during the period of the analysis; 
and iii) content is shown in the seeker's wall in chronological order. 
During the data collection period, Twitter users could toggle between two different kind of wall timelines: Top Tweets (algorithm-based) and Latest Tweets (time-based).
The first one uses the Twitter algorithm to shuffle posts in what it suggests as a better order. Even in this case, Twitter exposes people to relatively recent tweets \footnote{Full details on the timeline algorithm in
\url{https://help.twitter.com/en/using-twitter/twitter-timeline}.}. The latter instead shows friends' tweets in chronological order, and it is the mode we assume in our study since we do not have access to Twitter's algorithms. For the same reason, we assume all the tweets that appear in the wall come from the followers. \footnote{ Twitter may occasionally add tweets from users not followed \url{https://help.twitter.com/en/using-twitter/twitter-timeline}}

%% file: sec/04-dataset.tex
\section{Dataset}
\label{sec:dataset}

We use the Twitter API \footnote{\url{https://developer.twitter.com/en/docs}} to gather data from 1\textsuperscript{st} July 2020  until 11\textsuperscript{st} November 2020, enough time so that ambient awareness can be developed \cite{leonardi2015ambient}. %
In this time period, we collected activity of users (i.e, tweets, retweets, likes and replies/answers), following-follower relationships and List meta-data (i.e, creation time, creator, name, description and members). All this information allows us to i) identify users as seekers and/or experts alongside with their  topic of expertise, ii) extract interactions between seekers and experts, and iii) reconstruct the wall of the seekers and thus analyze exposure to content. In the following, we describe the procedures for gathering and filtering the data we use to explore IWAA (see Appendix \ref{app:dataset} for further details).

\begin{figure}[t]
\vspace{-0.3cm}
	\centering
	\includegraphics[width=0.99\linewidth]{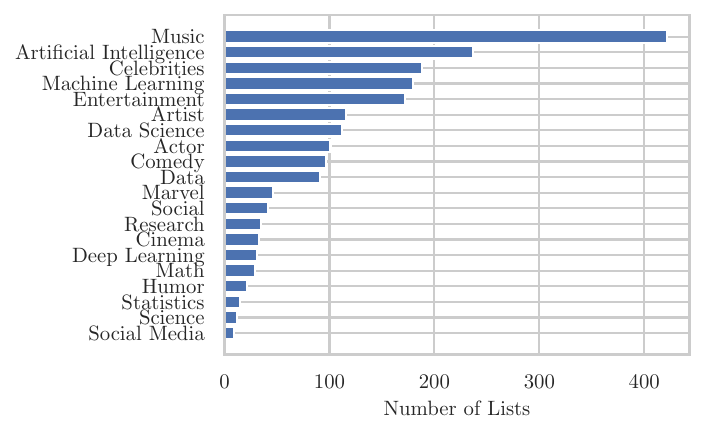}%
	\vspace{-0.3cm}
	\caption{Topics considered in our analysis and number of Lists collected for each of them.}%
	\label{fig:topics}
	\vspace{-0.3cm}
\end{figure}

\vspace{-0.3cm}
\paragraph{Data collection \& filtering} First of all, we select a number of topics of expertise, e.g. Artificial Intelligence and Statistics,  for which we are interested in analyzing IWAA. All of the topics considered are shown in Figure \ref{fig:topics}, together with the number of related Lists. 
In total we collected 2002 Lists, created  by  1836 seekers from 1\textsuperscript{st} August 2020  until 11\textsuperscript{st} November 2020.
Among the selected seekers, we filter out those for which their activity is private;  they are completely passive (i.e., their  timeline is empty, $|\activityp{s}|=0$, or their number of likes is zero, $|\activityl{s}|=0$); they do not follow any other user, or on the contrary, they follow over 5000 users---IWAA cannot be developed in an empty or too overcrowded wall. 
After the filtering, 1682 Lists and 1541 seekers remain for our study, which correspond to 84.02\% and 83.93\% of the initial values.
From these Lists, we discover a total of 12033 experts, from which only 16 happen to be also seekers. Thus, in total we have 13558 unique users and 36895 pairs seeker-expert.
Additionally, to facilitate the reproducibility of the results, we provide a Github repository (\url{https://github.com/XXX/XXX}) with the scripts used for data collection \& IWAA analysis as well as the identifiers of the Lists, users and tweets collected.

\begin{figure}[t]
\vspace{-0.3cm}
	\centering
	\includegraphics[width=0.55\linewidth]{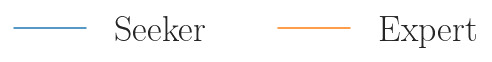}\\%
	\includegraphics[width=0.49\linewidth]{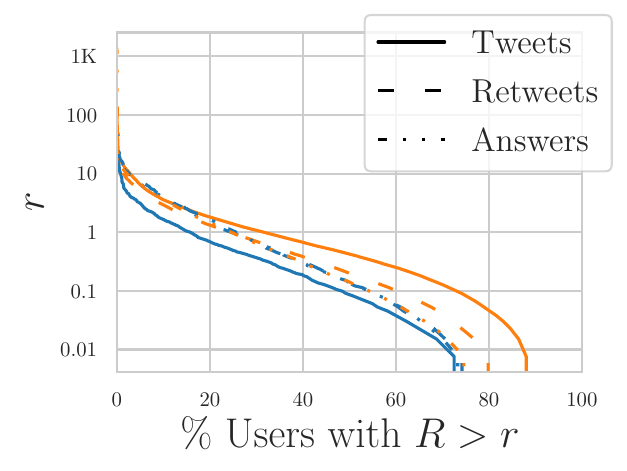}%
	\includegraphics[width=0.49\linewidth]{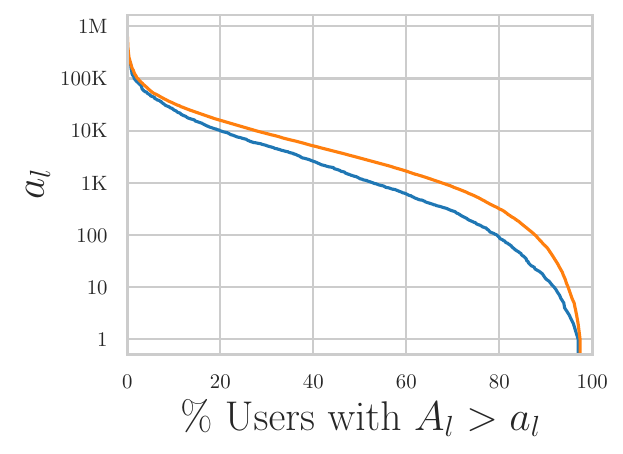}\\%
	\includegraphics[width=0.49\linewidth]{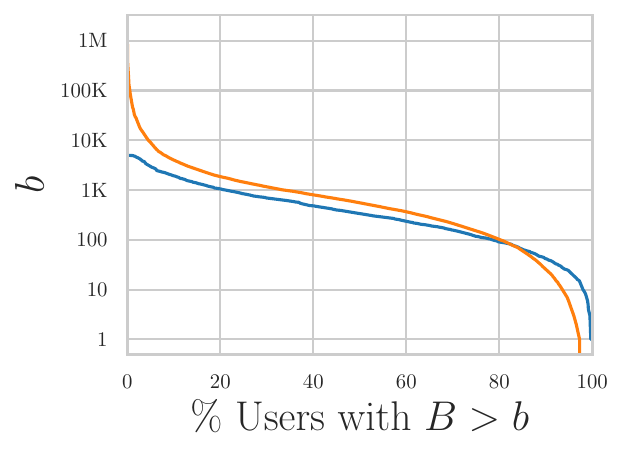}%
	\includegraphics[width=0.49\linewidth]{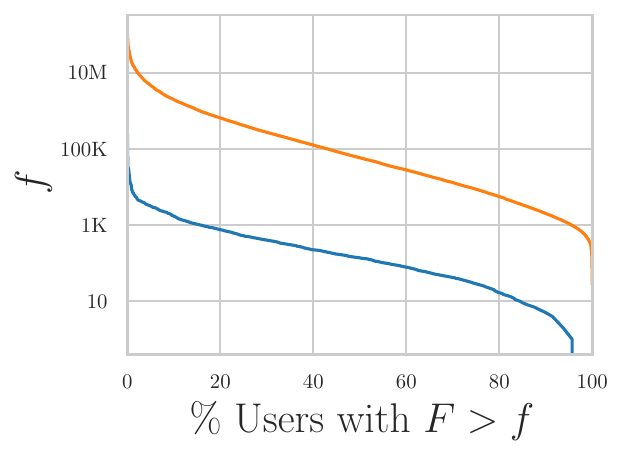}%
	\caption{We show the ICDF for  different features of the seekers (blue) and the experts (orange). Specifically, we show  the rate of posts $R$ (separated by tweets, retweets and answers), the count of likes $A_l$, friends $B$, and followers $F$.}%
	\vspace{-0.3cm}
	\label{fig:cdf_features_sb}
\end{figure}

\begin{figure*}
\vspace{-0.1cm}
	\centering
	\begin{minipage}[b]{.3\textwidth}
	\centering
$\vcenter{\hbox{\includegraphics[width=0.3\linewidth]{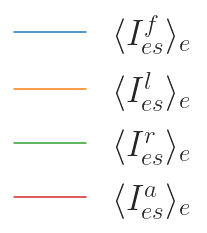}}}$%
$\vcenter{\hbox{\includegraphics[width=0.7\linewidth]{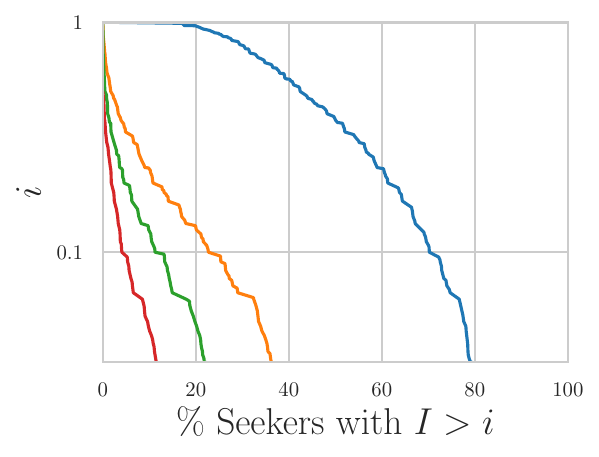}}}$%
\caption{ICDF for the four average reaction signals, we generally represent with $I$.\vspace{-0.5cm}}%
\label{fig:interactions_dataset}
	\end{minipage}\qquad
	\begin{minipage}[b]{.66\textwidth}
	\centering
\includegraphics[width=0.7\linewidth]{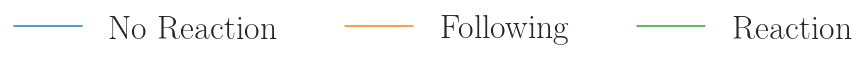}\\
\subcaptionbox{\label{fig:cluster_prop}}{\includegraphics[width=0.24\linewidth]{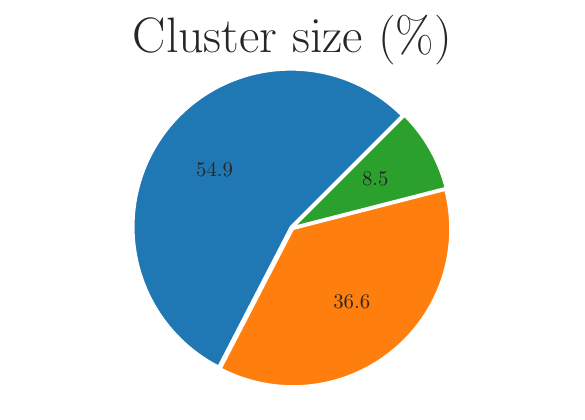}}
\subcaptionbox{\label{fig:cluster_avg_signals}}{\includegraphics[width=0.75\linewidth]{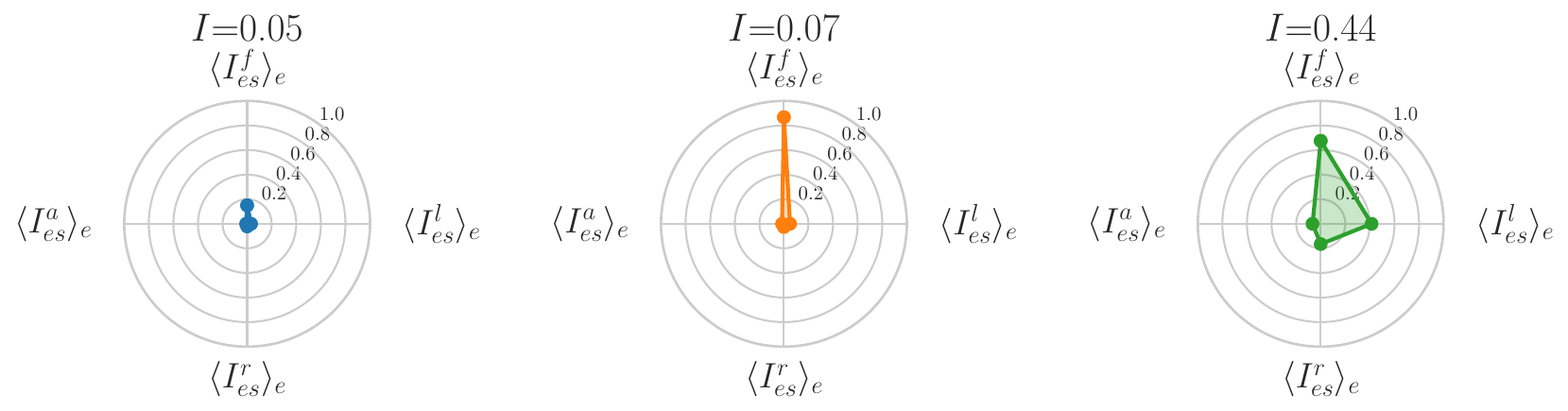}}
\caption{Clustering results. (a) Size proportion of each cluster. (b) Average interaction signals per cluster.\vspace{-0.5cm}}%
\mbox{}
\label{fig:clustering_avg_interactions}
	\end{minipage}
\end{figure*}

\subsection{Data exploration}

Prior to the IWAA analysis, we conduct an exploration of the collected data. In particular, we focus on characterizing the seekers and the experts according to some of their online features, e.g., number of friends. Figure \ref{fig:cdf_features_sb} shows, for the experts (in orange) and seekers (in blue), the inverse empirical cumulative distribution function (ICDF) of the rate of posts (divided by tweets, retweets and answers), the count of likes 
\footnote{Given that we do not need to collect the likes for the experts, we display the count of likes instead of the rate.}, 
friends and followers. 
The ICDF allows us to compare different populations: it reads as the percentage of the population larger than a certain value. This means it allows us to compare quantiles ---e.g., the median of the distribution: the slower a ICDF decreases, the larger the quantiles are.

Based on Figure~\ref{fig:cdf_features_sb}, we can highlight several points. First, the most clear difference between seekers and experts is in the number of followers:  experts are followed by many more people than seekers, whereas they do not differ so much in the number of followees. This is a first hint that experts are popular and enjoy a high reputation. %
Second, if we focus on the top left figure, we can see that experts generate and react to more content than seekers, i.e., they are more active users, and thus, more likely to be able to exhibit their expertise, which is a precondition for IWAA. This figure also reveals that experts create more tweets than replies or answers, whereas this is not the case for most the seekers. This result further strengthens our previous argument. As an additional remark, we discovered that around 20 \% of the seekers are listed in 10 or more Lists, which indicates they could be experts on some topics.

%% file: sec/05-iwaa_interaction.tex
\section{IWAA based on reactions}
\label{sec:iwaa_interaction}
In this section, we analyze the seekers' reactions to the~experts' content, as a strong evidence of the seeker's IWAA about an expert. First, we present several quantitative metrics for the reactions alongside with their interpretation. Then,~we use these measurements to group seekers with similar signal patterns and we relate each of them with an online behavior.

\subsection{Measuring the reactions}
\label{sec:interactions}

Twitter provides several ways in which the users can interact with each other. The interaction can be at network level via following other people, which we represent with $f$; or at content level via retweet, like and reply/answer, which we refer to as  $r$, $l$ and $a$ respectively. It is the content level interaction that provides evidence about the plausibility of IWAA. Concretely, the retweet and like since they require little cognitive effort---not even deliberate processing---from the user and and are driven from the content of the post. As a consequence, they constitute more meaningful signals of the existence of IWAA than the reply, which involves more complex processing and can be steered by the user's reputation \cite{cha2010measuring}. 
The follow interaction must be treated cautiously. It indicates the seeker has interest in receiving the content generated by the expert but it can also be related to bond-based relationships, which could mislead our analysis.
\vspace{-0.4cm}
\paragraph{Remark} In early 2020 Twitter launched the Quote Tweet (qtweet), which is just a retweet with some additional text. In our analysis, we consider qtweets as standard retweets. We observed that, in the data collected, qtweets represent only 20\% of the retweets, and  contain shorter text than answers---on average, answers contain 100.94 characters while it drastically drops to 67.17 for qtweets--- which indicates they require less cognitive effort to be created.
 
We are interested in the reactions from the seeker to  the expert, which we  represent with a binary variable $\Ix$ taking value 1 when there is interaction, and 0 otherwise. The superscript determines the type of interaction, i.e. $\text{x} \in \{r, l, a, f\}$. For example, $\IRT{es}=1$ indicates the seeker has retweeted at least one tweet of the expert (within the studied period). Then, from the point of view of a seeker, we can compute the percentage of experts that he reacts to as $\avg[e]{\Ix}$. We can also compute average percentage of reactions from seekers to experts, i.e.,  $\Ixs = \avg[e,s]{\Ix}$. 
Additionally, we refer to the total average of \textit{effortless reactions} as $\sumI{} =  \avg[e,s]{\IRT{es} \vee \Ilik{es}}$. 

Figure,~\ref{fig:interactions_dataset} shows the distribution for each type of reaction as a ICDF. Here, we observe that the percentage of seekers that react to experts' content is very small compared to the number of experts they follow, and the more cognitive capacity required for the reaction, the lower the \% of experts they interact with. For example, while $\sim70\%$ of the seekers follow at least 10\% of their experts, when looking at likes the percentage of seekers reacting to at least 10\% of their experts drops to $\sim20\%$  (and less than 10\% for  retweets and replies/answers).

\subsection{Clustering the seekers: passive behavior} 
\label{sec:clustering}
Here, we group the seekers according to their reactions, as measured above, and  explore what other features (e.g., number of posts)  tell us about seekers' awareness for different seeker groups.

\vspace{-0.4cm}
\paragraph{Experimental setup} We cluster seekers using our three types of interactions---that is $\avg[e]{\IRT{es}}$, $\avg[e]{\Ilik{es}}$ and $\avg[e]{\Irep{es}}$---as input features to the clustering algorithm. Notice that these metrics are already normalized between $[0,1]$, so no further pre-processing is needed. We cross-validate different algorithms and parameters configurations using the Silhouette coefficient \cite{rousseeuw1987silhouettes} as the quantity to be maximized. Spectral clustering \cite{von2007tutorial} happens to perform the best. See Appendix~\ref{app:clustering} for further details on the cross-validation scheme.

\begin{figure}[t]
\vspace{-0.62cm}
	\centering
	\subcaptionbox{Tweets rate\label{fig:cluster_e_tweets}}{\includegraphics[width=0.49\linewidth]{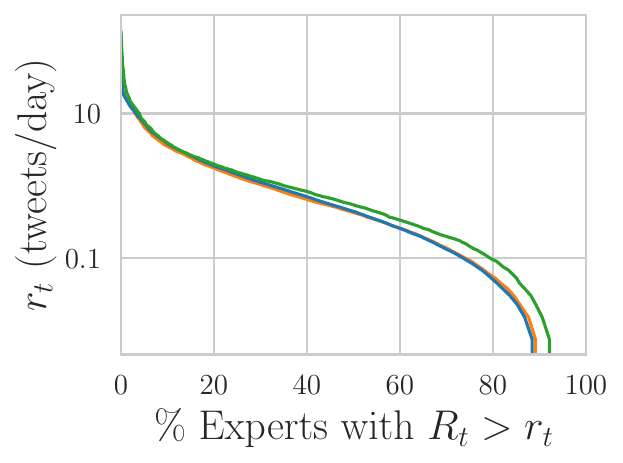}}
	\subcaptionbox{Retweets rate \label{fig:cluster_e_retweets}}{\includegraphics[width=0.49\linewidth]{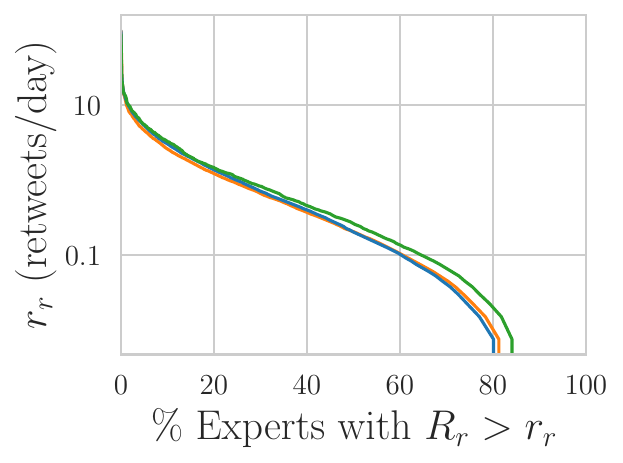}} \\
	\vspace{0.2cm}
	\subcaptionbox{Answers rate \label{fig:cluster_e_answers}}{\includegraphics[width=0.49\linewidth]{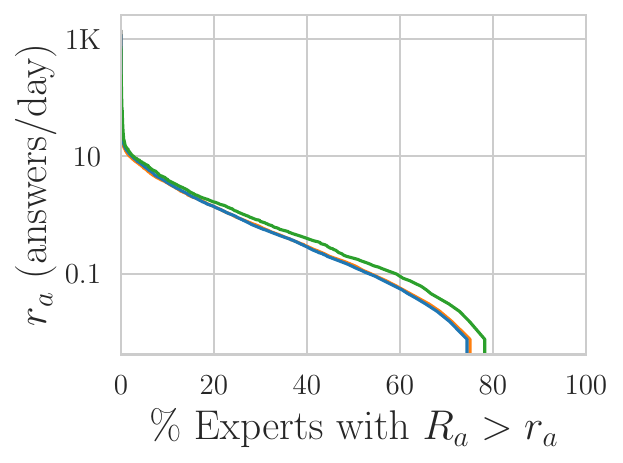}}
	\vspace{-0.2cm}
	\caption{ICDF of experts' features per cluster.}%
	\vspace{-0.4cm}
	\label{fig:clustering_expert_features}
\end{figure}

Figure \ref{fig:clustering_avg_interactions} summarizes the clustering results. Specifically, Figure \ref{fig:cluster_avg_signals} shows the four  average reaction metrics for the  clusters, which are ordered  from lower (on the left) to higher (on the right) level of  effortless reactions measured as $\sumI{} =  \avg[e,s]{\IRT{es} \vee \Ilik{es}}$. We also show the proportion of users on each cluster in Figure \ref{fig:cluster_prop}.
Here, we observe that over 90\% of the seekers do not interact at all with their experts (blue and orange clusters). This matches again our prior intuition on the passive behavior of Twitter users and the idea that ambient awareness does not require interaction, but may arise just from skimming updates. In contrast, the seekers belonging to the third cluster (respectively, 8.5\% of the seekers), present higher level of interactions to their experts. 
We accompany these results with several features of the experts (see Figure \ref{fig:clustering_expert_features}) that the seekers in each cluster interact with. Notice that the cluster with more seekers' reactions is in turn also characterized  with experts presenting by high retweet and answer ratio.

\begin{figure*}[t]
    \centering
    \includegraphics[width=0.5\linewidth]{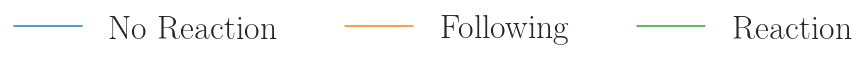}\\
\end{figure*}

Lastly, in Figure \ref{fig:clustering_seeker_features} we display the features of the seekers in each cluster. Notice that the cluster with more reactions, the green one, comprises the most active seekers in all types of posts: they exhibit the highest rates of tweets, retweets, answers and likes.
As a last remark, we also explored the percentage of seekers that are  potential experts, i.e., listed 10 times or more. As shown in Appendix \ref{app:clustering}, we observed that the distribution remains similar across clusters, which does not provide information about the composition of the cluster, e.g., the existence of an "only experts" group.

From this empirical evidence, we can conclude that the rate of activity of the seekers relates with the amount of reactions to the experts: the most active a seeker is, the most likely he interacts with experts. Unsurprisingly, a large percentage of the seekers---around 90\%---do not react to the experts' content. This passiveness of the seekers motivates us to introduce the concept of in-wall visibility. We thus continue with the analysis of IWAA from a perspective that does not need explicit actions from the user, i.e., we focus on passive exposure of the content of the experts.

%% file: sec/06-iwaa_visibility.tex
\section{IWAA beyond interactions: Exposure to content}
\label{sec:iwaa_visibility}

In this section we further investigate the limits of IWAA by looking at the visibility of the expert's content in the seeker's wall prior to the creation of the List, as it does not require any further effort from the seekers' side than skimming through the wall. In  particular, we provide a framework to quantify the time that the content of an expert $e$ is exposed (or visible) in the wall of a seeker $s$, which we refer to as $\Vis$, and then use this framework to study the plausibility of developing IWAA.

\vspace{-0.4cm}
\paragraph{Notation} Let  $\activity{es}$ be the sequence of posts of the expert that appear in the seeker's wall, and $t_l^e$ the time of the last post in $\activity{es}(t)$.
Let {$\wall{s/e}$$=$$\wall{s}$-$\activity{es}$} be the sequence of posts in the wall of the seeker excluding the ones from the expert.

\begin{figure}[t]
\vspace{-0.57cm}
	\centering
	\subcaptionbox{Tweets rate\label{fig:cluster_s_tweets}}{\includegraphics[width=0.49\linewidth]{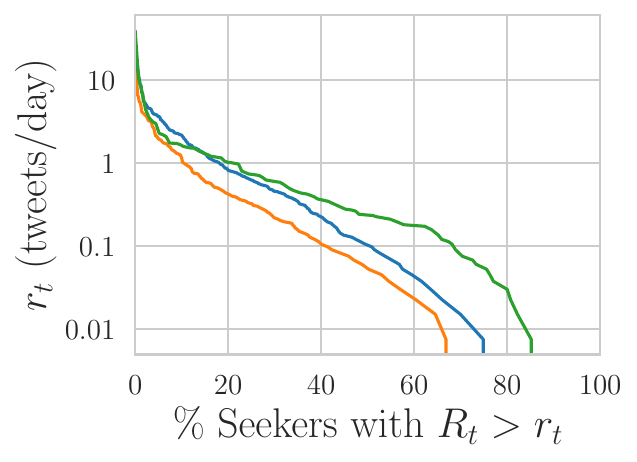}}
	\subcaptionbox{Retweets rate\label{fig:cluster_s_retweets}}{\includegraphics[width=0.49\linewidth]{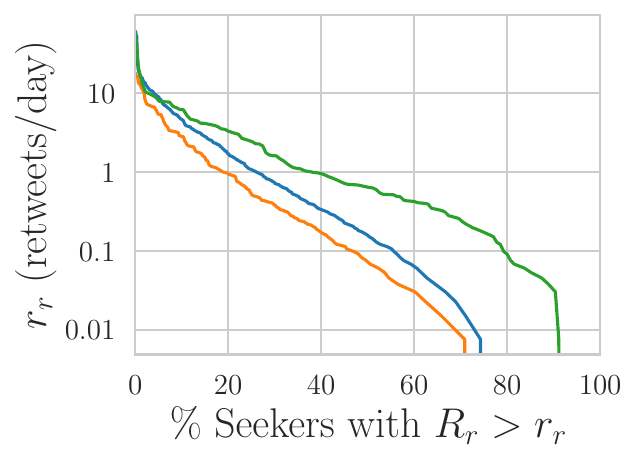}} \\ \vspace{5pt}
	\subcaptionbox{Answers rate\label{fig:cluster_s_answers}}{\includegraphics[width=0.49\linewidth]{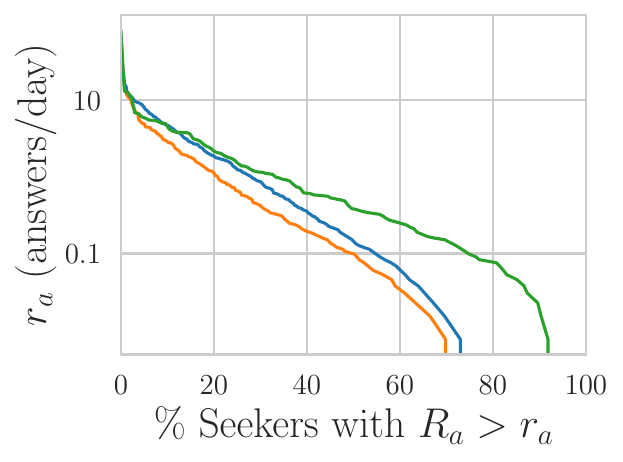}}
	\subcaptionbox{Likes rate\label{fig:cluster_s_likes}}{\includegraphics[width=0.49\linewidth]{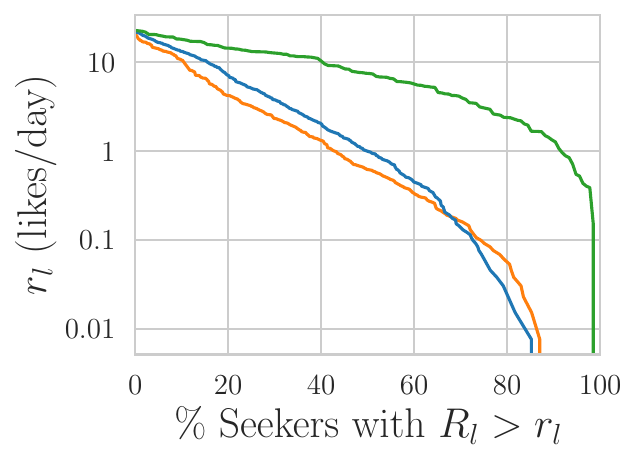}}
	\vspace{-0.2cm}
	\caption{ICDF of seekers' features per cluster.}%
	\vspace{-0.2cm}
	\label{fig:clustering_seeker_features}
\end{figure}

\subsection{Bounds for the visibility}
\label{sec:visibility}
We propose to compute the visibility of the content of an expert $e$ in the wall of a seeker $s$, or in-wall visibility, over the interval of time $[t_1, t_2]$  solving the integral:
{\small
\begin{align}
\Vis(t_1, t_2)&:=\int_{t_1}^{t_2} f_{es}(\tau)  \pres[s]{\tau}  d \tau,
\label{eq:agg_vis}
\end{align}
}
where the function $f_{es}(t)$ corresponds to the probability of the content of $e$ being visible in the wall of $s$ at time $t$, and $\pres[s]{t}$ refers to the probability of the seeker being present in the wall, i.e., online,  at time $t$. In the following, we omit the interval limits to improve the overall readability. Intuitively, the higher $\Vis$, the more likely the seeker becomes aware of the expertise of the expert while skimming through the wall. Importantly, if $\Vis=0$, then IWAA is impossible and thus  awareness of experts must come from somewhere else but the Twitter wall, potentially even from offline activity. Next, we detail how to compute in-wall visibility, specifically, how to model both functions,  $f_{es}(t)$ and $\pres[s]{t}$.

\vspace{-0.4cm}
\paragraph{Modeling exposure} The probability of exposure of a post of $e$ in the wall of $s$ ($f_{es}(t)$) should be high when an expert's post appears at the top of the timeline. Then, as we assume a chronological wall, it decreases as posts from other friends arrive to the wall, making on average more likely that the seeker will not reach/see it. Following this intuitive idea, we propose the following modeling:
{\small
\begin{equation}
f_{es}(t;k, m) =  f_{es}(t) =  \left[1 - \frac{\min(n_{es}(t), k)}{k} \right]^m,
\label{eq:ins_vis}
\end{equation}
}
where $n_{es}(t) = |\wall{s/e}(t_l^e, t)|$ is the number of posts in the wall of $s$ since the last post of $e$, the parameter $k$ defines up to which number of events in the wall $f_{es}(t;k, m)$ is non-zero, and the parameter $m$ determines the decay of the function. Figure~\ref{fig:ins_vis} depicts the shape of this function for $n_{es}(t) \in \{0,...,100\}$ for different configurations of the parameters $k$ and $m$.  Intuitively, both parameters should be chosen according to the distribution of the number of posts the seeker skims in every session of activity: $k$ is related to the maximum number of posts while $m$ relates to the mean.

\vspace{-0.4cm}
\paragraph{Modeling online presence} We model the probability of an user $s$ being  online at time $t$,i.e., $\pres[s]{t}$,  as an asymmetric Laplace kernel centered at the time of the closest post $t_c$:
\begin{equation}
\pres[s]{t} =   e^{- \frac{||t - t_c||}{ a_l \left[ t \leq t_c \right] +  a_r \left[ t > t_c \right]}  } \text{ with }
t_c = \argmin_{t_j \in \activityp{s} } d(t_j, t),
\label{eq:online_presence}
\end{equation}
where the parameters $a_l$ and $a_r$ determine how the kernel decays to the left and to the right, respectively. These parameters can be chosen globally or locally, i.e., per event, and intuitively capture how the probability of being online before and after a post decreases as we get further away from the time of the post. 
Specifically, we assume here that $a_l=a_r=0.047$ (hours) for all the events, but for those consecutive posts that are closer than the average length of a session---i.e., $\approx 4$ minutes according to Statista (see further details below in the hyperparameter selection), in which interval we assume the seeker to be continuously active.
As shown in~\cite{kooti2016twitter}, Twitter users tend to show activity at the beginning and the end of a session, a behavior we could capture with the  parameters $a_l$ and $a_r$.
This simple but rather flexible modeling, allows to capture online user behavior with a small number of parameters. 
Refer to Appendix \ref{app:visibility} for an illustrative example.

Unfortunately, as mentioned previously, most of Twitter's users are passive. This means, with our modeling choice of the online presence, we  miss any inactive session. Moreover, we cannot consider the like activity of the user because Twitter only provides the creation time of the post but not the "like" time. 
As a consequence, with the modeling of the online presence $\pres[s]{t}$ described above,  we only obtain a loose lower bound (LB) on the true online presence, and hence, we are only computing a LB on the true visibility $\Vis$. We thus refer to this LB as $\VisLB$. %
If alternatively, we consider the user $s$ is always online, that is $\pres[s]{t}=1$, we obtain an upper bound (UB) of the visibility $\Vis$, which we denote $\VisUB$. Now, we know the true unavailable visibility  $ \Vis$ lays in $\VisLB < \Vis < \VisUB$.

As remarks, notice both $f_{es}(t)$ and $\pres[s]{t}$ are piecewise functions. This enables to solve the integral in Equation \ref{eq:agg_vis} as a sum over the elements in $\wall{s} \cup \activityp{s}$. Also, notice that the visibility is independent on the topic of expertise, it just depends on the pair seeker-expert.

\subsection{Analysis of the visibility}

In this section we study visibility/exposure for the pairs seeker-experts introduced in previous sections, for which we first need to reconstruct each seeker's wall. 
As recovering each seeker's wall requires retrieving all the posts of his friends over the considered time interval, we here focus the study on a random uniform sample of 137 seekers from the cluster without reactions (blue and orange) found in Section \ref{sec:iwaa_interaction} that represents 90\% of the seekers. To be precise, we restrict our analysis to the 123 seekers with less than 2500 friends: we observed the average visibility decays quickly as we increase the number of friends (see Appendix \ref{app:visibility}).
We compute the visibility per day for the 30 days prior to the creation of the \List. Then, we take the average over the days and the experts to obtain the average visibility  for a seeker, i.e., $\avg[\text{day, e}]{\Vis}$.  
We remark that  $\avg[\text{day, e}]{\Vis}$ can be interpreted as the expected time (per day) that the seeker $s$ is exposed to the content of his experts, i.e., the average visibility of the experts in the seeker's wall. Also, we remind that in the following we only study the experts that the seeker has included in his List. 
We split the analysis in two parts by considering two disjoint set of experts: the ones that are followed by the seekers, we refer to as $\Ecal_f$, and the ones that are not followed by, that is  $\Ecal$. %
Additionally, we have repeated the analysis below on a wall resulting from filtering the least popular Tweets as a proxy to the unavailable Top Tweets Wall provided by Twitter. The conclusions reached are similar.

\vspace{-0.4cm}
\paragraph{Hyperparameters' selection} We analyze the sensitivity of the results to the choice of the  hyperparameters $k$, $m$ of the exposure function $f_{es}(t)$. As a starting point to delimit the range of values, we rely on previous work that characterizes users' behavior in OSNs. In particular, the online portal Statista revealed in its report of September 2019 that Twitter users stay  on average less than $4$ minutes per session on the platform \footnote{https://www.statista.com/statistics/579411/top-us-social-networking-apps-ranked-by-session-length/}.

\begin{figure*}[t]
\vspace{-0.1cm}
	\centering
	\includegraphics[align=t,width=0.14\linewidth]{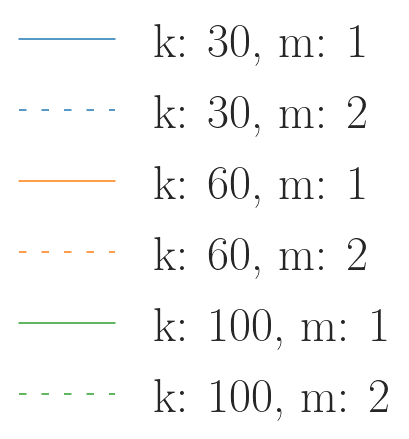}%
	\includegraphics[align=t,width=0.25\linewidth]{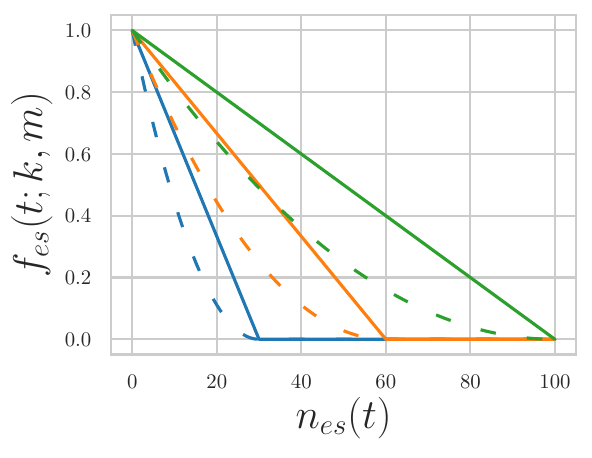}%
	\includegraphics[align=t,width=0.26\linewidth]{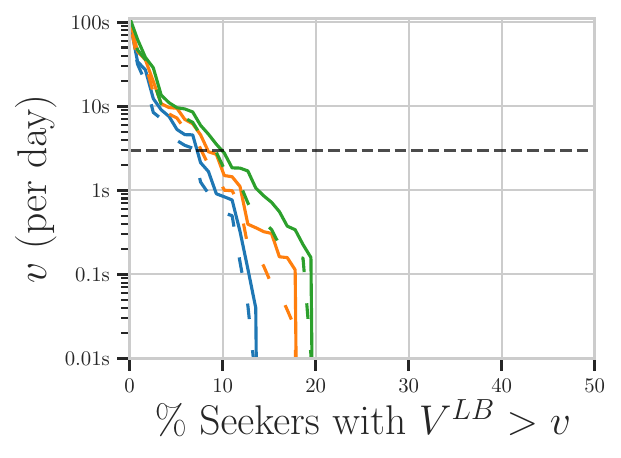}%
	\includegraphics[align=t,width=0.26\linewidth]{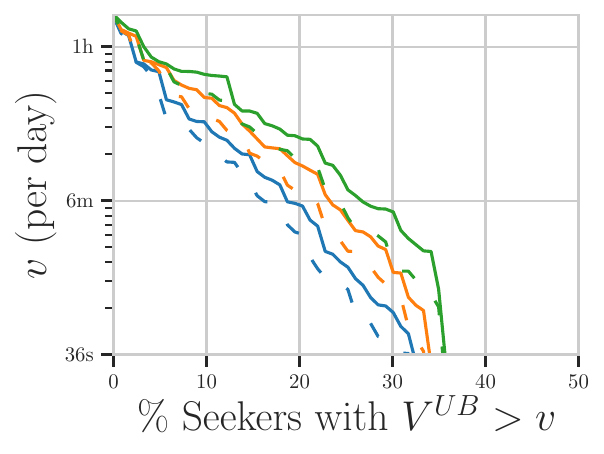}%
	\vspace{-0.2cm}
	\caption{On the left, plot of $f_{es}(t;k, m)$ as $n_{es}(t)$ increases for different values of the parameters $k$ and $m$. On the two right-most figures, ICDF of  the lower  $V^{LB}$ and upper $V^{UB}$ bounds of the visibility for different configurations of $k$ and $m$.}%
	\label{fig:ins_vis}
	\vspace{-0.4cm}
\end{figure*}

About the time it takes to skim a tweet, \citet{counts2011taking} observed users only allocate a couple of seconds to read each tweet. This time frame also matches the setup up of spontaneous inference studies \cite{levordashka2017spontaneous, todorov2003efficiency}, where users are exposed to messages between 2 and 5 seconds. Then, we assume that it takes $\approx 3$ seconds to skim through a post and  consider $k \in \{30, 60, 100\}$ and $m \in \{1, 2\}$. Figure \ref{fig:ins_vis} (two right-most plots) shows the ICDF for both bounds of the average visibility over the experts for different configurations of the parameters $k$ and $m$. 
We observe that, while as expected the exposure increases as we increase $k$ or decrease $m$, the trace of the visibility is consistent across settings. For a given time of average exposure, the maximum deviation in  the percentage of seekers  is 10\%, which happens when $V^{UB} \approx 6$ minutes. Appendix \ref{app:visibility} contains further details on the hyperparameter selection for the online presence function.  
For a deeper analysis of the visibility, we compute the bound with $k$=$100$ and $m$=$2$. Also, we divide the following analysis into two groups of experts, namely $\Ecal_f$ and $\Ecal$.

\begin{figure}[t]
	\centering
	\includegraphics[width=0.75\linewidth]{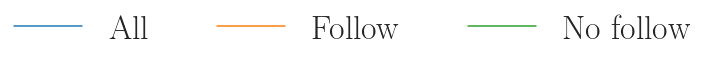}\\%
	\includegraphics[width=0.49\linewidth]{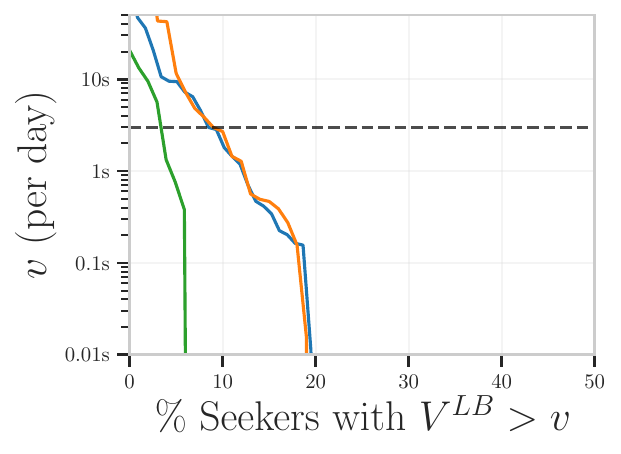}%
	\includegraphics[width=0.49\linewidth]{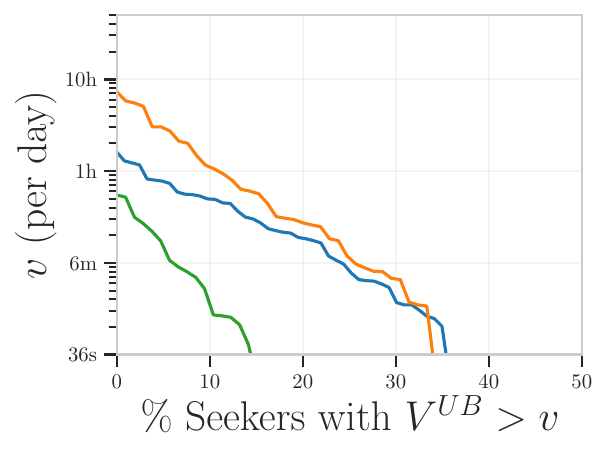}%
	\vspace{-0.2cm}
	\caption{ICDF of the bounds of the visibility averaged over experts and days. In blue the average is over all the expert; in orange over $\Ecal_f$; and in green over $\Ecal$. On the left we show $\avg[e, day]{\VisLB}$ with a dotted black line at a visibility level of 3 seconds. On the right we show $\avg[e, day]{\VisUB}$.}%
	\vspace{-0.2cm}
	\label{fig:visibility_ub_lb_cdf}
\end{figure}

\vspace{-0.4cm}
\paragraph{What does the lower bound tell us? } 
Figure~\ref{fig:visibility_ub_lb_cdf} on the left shows the ICDF for the average LB of the visibility per day for the seekers---that is $\avg[e, day]{\VisLB}$ or $\VisLBs$ in short. In particular, we display the $\VisLBs$ for $\Ecal_f$ in orange, for $\Ecal$ in green, and for the whole population of experts in blue. There are several important points to note. 
First, notice that all the values found for $\VisLBs$ lay in the order of seconds per day or less. Even though they may seem meaningless, a percentage of them refers to one tweet per day of exposure on the 30 days previous to the List creation time, which in fact seems a reasonable amount of exposure to develop IWAA. By choosing 3 seconds per day (black dashed line) as evidence for the possibility of developing IWAA, then we observe that $11\%$ of the seekers show strong evidence on the possibility to develop IWAA for the experts that they are already following, i.e., for the experts in $\Ecal_f$. 
This percentage is surprisingly high taking into consideration we have a loose lower bound on the true visibility. 
On the other hand, this percentage significantly drops to $3.92\%$ for the experts that are not followed by the seekers. This finding tell us that it is significantly easier for the seeker to develop IWAA if he follows the expert. However, in order for the seeker to follow an expert,  either Twitter has recommended the expert as a potential connection, he knows the expert offline, or he must have being exposed to the expert via the retweets of his friends. 

\vspace{-0.4cm}
\paragraph{What does the upper bound tell us? } Figure~\ref{fig:visibility_ub_lb_cdf} on the right shows the ICDF for the averaged upper bound visibility, namely $\avg[e, day]{\VisUB}$ or $\VisUB$ in short. As before, it is also splitted by the two experts' groups. The first thing we notice is the difference in time-scale compared to the visibility lower bound, as $\VisUB$ reaches up to 10 hours.  
This already tells us we are dealing also with a loose upper bound. Yet, we observe that for more than 60\% of the seekers the exposure to the experts' content in $\Ecal$ is less than 36 seconds, the exposure for this group of experts always being below 30 minutes. 
This indicates that \emph{seekers are unlikely--- if it is even possible---to develop IWAA for experts they do not follow}. Notice we are considering that the seeker is always online when in reality, on average, only 40\% of the Twitter users connect daily for less than 4 minutes, which would need in turn to overlap with the less than  36 seconds (of 30 minutes in the best case) that the experts are visible in his wall. 
What is even more striking is that for the overall population of experts, around 60\% of the seekers present zero exposure to the experts' content, even when considering a visibility upper bound.  Thus, it seems that \emph{for at least 60\% of passive seekers, which are in turn the majority (over 90\%), IWAA is impossible}, and then the awareness should come from a different use of Twitter (e.g., the Explore timeline) or alternatively from offline activity.

%% file: sec/07-discussion.tex
\begin{figure*}[t!]
\vspace{-0.25cm}
	\centering
	\begin{minipage}[b]{.3\textwidth}
	\centering
	\includegraphics[width=0.99\linewidth]{images/clustering_legend.png}\\%
\includegraphics[width=0.76\linewidth]{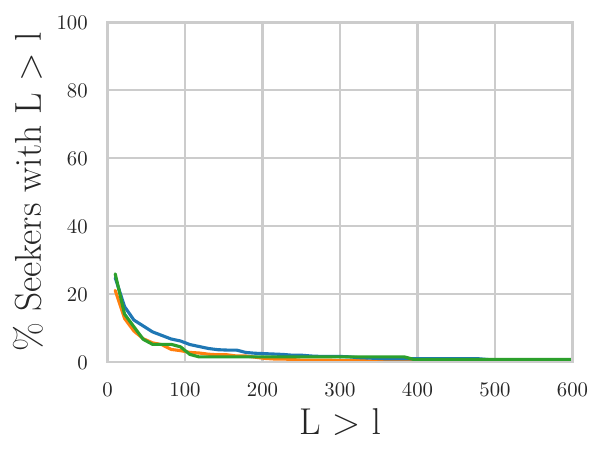}%
\caption{Plot of the percentage of seekers listed $l$ times or more in the each of the three clusters.}%
\label{fig:app_seekers_expertise}%
	\end{minipage}\qquad
	\begin{minipage}[b]{.3\textwidth}
	\centering
		\includegraphics[width=0.9\linewidth]{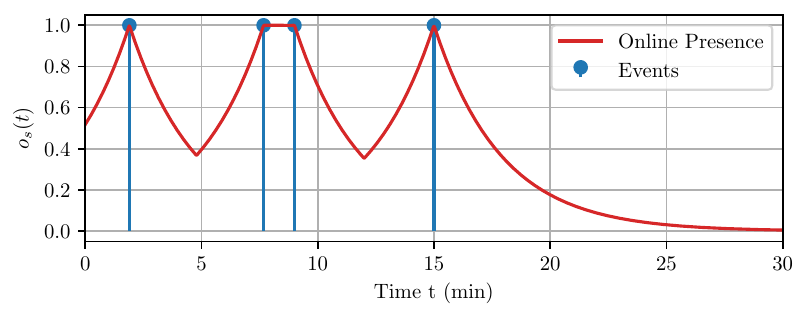}%
	\caption{Illustrative example of the probability of online presence for a seeker with four events, i.e., posts in twitter. }%
	\label{fig:online_presence_example}
	\end{minipage}\qquad
		\begin{minipage}[b]{.3\textwidth}
	\centering
		\includegraphics[width=0.9\linewidth]{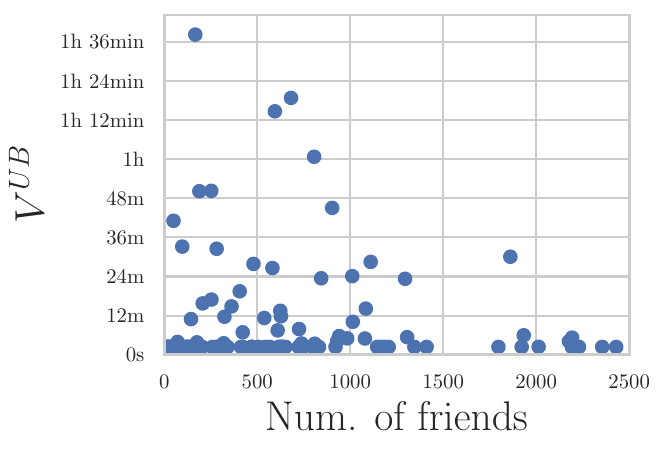}%
	\caption{Average upper bound visibility per day vs number of friends of the seekers for pairs under study. }%
	\label{fig:vis_vs_friends}
	\end{minipage}
\end{figure*}

\section{Discussion}
\label{sec:conclusion}

Until now, ambient awareness---the knowledge about who-knows-what by simply skimming through online updates--- was explored via user studies, which do not easily scale up to the network size, as they are costly and rely on subjective self-report data, e.g., on the answers of the participants.
Up to the best of our knowledge,  this paper is pioneer at developing a  data-driven methodology to study the boundaries of ambient awareness in the wall of Twitter (IWAA).
The result  of our analysis are in line with \citet{levordashka2016ambient} who found that most Twitter users reported ambient awareness only for some of their network members. Yet, they did not know whether this was due to the limited cognitive capacity of the users or the fact that many Twitter users do not tweet  enough or do not post tweets that are diagnostic of their expertise. The present results indicate that it might be the low frequency and/or diagnosticity of the tweets in a user's wall that limits ambient awareness. Our data-driven approach nicely complements user studies and presents an important contribution to work on ambient awareness. 

We remark that our analysis considers a significant number of Twitter users (36895 pairs seeker-expert), focuses on  diverse topics such as Social Science or Math, and  is robust to the hyperparameter choice. As a consequence, 
we expect that the main results from this analysis are not sensitive to the considered dataset--the specific percentages of the bounds may slightly vary across datasets but not the main insights. %

We would like to highlight that the presented data-driven methodology paves the way for further  research studies of ambient awareness at a large scale. For example, to explore the boundaries of IWAA in different topics and communities of Twitter.
Moreover, data-driven approaches allow  studying more fine-grained queries, such as: Is IWAA a property of the seeker? If so, what are the user characteristics that make him/her suitable to develop IWAA? Does IWAA depends on the experts a user follows? To answer these and other similar queries, one would need to perform a detailed temporal analysis of  users' wall, which can be only performed efficiently by relying on data analysis tools, like machine learning.

%% file: appendix/app_dataset.tex
\section{Dataset details}
\label{app:dataset}
In order to start the collection of the Lists for the study, we selected several experts a priori that we know are active on Twitter, as shown in Table \ref{table:known_experts}. Then, we scrape the Lists in which they are members and continue the search for new experts among the members of the Lists.

\begin{table}[h!]
	\centering
	\small
	\begin{tabular}{c|l}
		\hline\hline
		Topic & Usernames  \\
		\hline \hline
		Music & ChiliPeppers, johnlennon, jonbonjovi, xtina \\
		\hline
		Entertainment & AnnaKendrick47, EmmaWatson, LeoDiCaprio,  \\
		& RobertDowneyJr, SamuelLJackson, TheRock,  \\
		&  VancityReynolds  \\
		\hline
		Artificial  & \_beenkim, drfeifei, fhuszar, goodfellow\_ian,  \\
		intelligence &   karpathy,mrtz, OriolVinyalsML,  \\
		 & RandomlyWalking, SchmidhuberAI, shakir\_za,  \\
		 &  yeewhye, ylecun, yudapearl, ZoubinGhahrama1  \\
		\hline\hline
	\end{tabular}
	\caption{Topical experts selected a priori that are active in Twitter.}
	\label{table:known_experts}
\end{table}

%% file: appendix/app_clustering.tex
\section{Clustering details}
\label{app:clustering}
We explore different algorithms and hyperparameter configurations to group the seekers according to their reactions to the content of the experts. In particular, we have cross-validated i)  $k\in\{3,...,10\}$ for KMeans, ii) $k\in\{3,...,10\}$ , $\gamma \in \{0.8, 0.6, 0.5, 0.4\}$ for Spectral clustering, iii) damping factor $\in \{0.6, 0.7, 0.8, 0.9\}$ for Affinity propagation, and iv) bandwidth$\in \{1, 0.8, 0.6, 0.5\}$ for Mean shift. Spectral clustering results to be the best choice with  $\gamma=0.8$ and $k=3$ and a Silhoutte Coefficient of 0.57. As remark, the Silhouette Coefficient quantify the affinity among points in the same cluster with respect to the other clusters.

\vspace{-0.3cm}
\paragraph{Seekers' expertise} We further explore the seekers' expertise regarding the analysis in Section \ref{sec:clustering}. Particularly, we pose the following  question: \emph{how many seekers in the different clusters are potential experts with expertise level $l$, i.e., are listed more than $l$ times?} Figure \ref{fig:app_seekers_expertise} shows the percentage of seekers  that are potential experts for different levels of expertise, i.e.  $L>l$, with $l \in [10, 600]$, for the three clusters.  The main thing we observe is that the distribution is quite uniform across clusters and the percentage of seeker decreases, regardless of the cluster, as we increse the minimum number of listed times, i.e., $l$. For example, for $l=10$  around 20\% of the seekers are potential experts independently on the cluster they lay on. This percentage drops close to 0 as we consider $l > 200$. This results tell that any of the clusters represent an "only experts" group, regardless of the minimum level of expertise considered.

\vspace{-.3cm}

%% file: appendix/app_visibility.tex
\section{Visibility details}
\label{app:visibility}
Figure~\ref{fig:online_presence_example} illustrates the shape of $\pres[s]{t}$, with the configuration of hyperparameters we use in the analysis. 
Specifically, we assume here that both  $a_l$ and $a_r$ are equal for all the events, but for those consecutive posts that are closer than the average length of a session---i.e., $\approx 4$ minutes according to Statista, in which interval we assume the seeker to be continuously active. This happens for the events between minutes 5 and 10 and allow us to model time intervals where we consider the user is online. 
In general, this design choice allows great flexibility to model online user behaviour, always with a trade-off between complexity ---i.e., number of parameters--- and simplicity. Then, for events further away than 4 minutes, we have the following expression $\pres[s]{t} =   e^{- \frac{||t - t_c||}{ a }  }$
 with $t_c = \argmin_{t_j \in \activityp{s} } d(t_j, t)$,
which we can also express in terms of the interval since the last event, i.e. $\Delta t =||t - t_c||$. We base the choice of the bandwidth of the kernel, i.e. $a$, using previous studies of online user behaviour in Twitter---recall from the main manuscript that the average session of a user lasts less than 4 minutes. 
Specifically, we select $a=0.047$ such that the probability of being online after 4 minutes  since the last event is  smaller or equal to 0.5, that is $o(\Delta t= 4min) < 0.5$. This also means that $o(\Delta t= 10min) = 0.031$, which makes sense.  In general, we can obtain a value for $a$ with  $ o(\Delta t) < c \rightarrow  e^{- \frac{\Delta t}{ a }  }  < c \rightarrow    a < \frac{\Delta t}{ - \log c}$. 

\vspace{-0.3cm}
\paragraph{Do the number of friends affect the visibility?} Figure \ref{fig:vis_vs_friends} show the scatter plot of the average upper bound visibility per day $\avg[day]{\VisUB}$ versus the number of friends of the seeker, i.e. $|\friends{s}|$. Notice that the visibility decreases with the number of friends, and for seekers with more than 2000 friends,it is quite low, never surpassing 12 minutes per day on average (right part of the plot).

\newpage